\documentclass[10pt]{article}
\usepackage{times}
\usepackage{mathfont}

\def\:{\kern 0.1em}

\usepackage{url}
\input epsf
\def\efig#1#2{\hbox{\epsfxsize=#1\epsfbox{#2}}}

\newtheorem{lemma}{Lemma}
\newtheorem{corollary}{Corollary}
\newtheorem{theorem}{Theorem}

\newcommand{\qed}{~$\Box$\medbreak}
\newenvironment{proof}{\noindent{\bf Proof: }}{\qed\medskip}

\def\log{\mathop{\rm log}}
\def\inf{\mathop{\rm inf}}
\def\dim{\mathop{\rm dim}}
\def\max{\mathop{\rm max}}
\def\exp{\mathop{\rm exp}}
\def\tan{\mathop{\rm tan}}

\DeclareSymbolFont{AMSb}{U}{msb}{m}{n}
\DeclareSymbolFontAlphabet{\Bbb}{AMSb}
\def\R{\ensuremath{\Bbb R}}
\DeclareMathSymbol{\subsetneq}{\mathrel}{AMSb}{"28}

\DeclareSymbolFont{lasy}{U}{lasy}{m}{n}
\let\Box\undefined
\DeclareMathSymbol\Box{0}{lasy}{"32}

\def\union{\cup}
\makeatletter

\long\def\@makecaption#1#2{
   \vskip 10pt 
   \setbox\@tempboxa\hbox{{\small #1. #2}}
   \ifdim \wd\@tempboxa >\hsize   % IF longer than one line:
       {\small #1. #2}\par        %   THEN set as ordinary paragraph.
     \else                        %   ELSE  center.
       \hbox to\hsize{\hfil\box\@tempboxa\hfil}  
   \fi}

\def\@begintheorem#1#2{\it\trivlist
				\item[\hskip \labelsep{\bf #1\ #2.\ }]}
\def\@opargbegintheorem#1#2#3{\it\trivlist
				\item[\hskip \labelsep{\bf #1\ #2\ {\rm(#3)}.}]}

\makeatother

\begin{document}

\title{Optimal Point Placement for Mesh Smoothing}

\urldef{\nahome}\url{http://www.geom.umn.edu/~nina/}
\urldef{\namail}\url{amenta@parc.xerox.com}
\urldef{\mbhome}\url{http://www.parc.xerox.com/csl/members/bern/}
\urldef{\mbmail}\url{bern@parc.xerox.com}
\urldef{\dehome}\url{http://www.ics.uci.edu/~eppstein/}
\urldef{\demail}\url{eppstein@ics.uci.edu}

\author{\ Nina Amenta~\thanks{Xerox Palo Alto Research Center, 3333
Coyote Hill Road, Palo Alto, CA, 94304; \nahome; \namail.}
\and
\ \ Marshall Bern~\thanks{Xerox Palo Alto Research Center, 3333 Coyote
Hill Road, Palo Alto, CA, 94304; \mbhome; \mbmail.  Work performed
in part while visiting Carnegie-Mellon University.}
\and
\ David Eppstein~\thanks{Department of
Information and Computer Science, University of California, Irvine, CA
92697-3425; \dehome; \demail. Work supported in part by NSF grant
CCR-9258355 and by matching funds from Xerox Corp, and performed in
part while visiting Xerox PARC.}}

\date{}
\maketitle

\begin{abstract}
We study the problem of moving a vertex in an unstructured mesh of
triangular, quadrilateral, or tetrahedral elements to optimize the shapes
of adjacent elements. We show that many such problems can be solved in
linear time using generalized linear programming. We also give efficient
algorithms for some mesh smoothing problems that do not fit into the
generalized linear programming paradigm.
\end{abstract} 

\section{Introduction}

Unstructured mesh generation, a key step in the finite element method,
can be divided into two stages.  In {\em point placement},
the input domain is augmented by {\em Steiner points}
(vertices other than those of the original domain) and a
preliminary mesh is formed, typically by Delaunay
triangulation.
In {\em mesh improvement}, local optimizations
are performed, involving the movement of Steiner points and
rearrangement of the mesh topology.

Computational geometry has made some inroads into point placement, and
methods including Delaunay refinement, quadtrees, and circle
packing are now known to generate meshes with guaranteed
quality; for surveys of these results, see~\cite{BE,BP}.
There has been less theoretical progress, however, in mesh
improvement, which has remained largely the domain of practitioners.

Mesh improvement typically combines several
kinds of local optimization:
\begin{itemize}
\item {\it Refinement and derefinement} split and merge triangles,
changing the number of Steiner points.
\item {\it Topological changes} such as {\em flipping} replace sets of
elements by other such sets, while preserving the positions of the
Steiner points.
\item {\it Mesh smoothing} moves the Steiner points of the mesh
while preserving its overall topology.
\end{itemize}

In this paper we study mesh smoothing algorithms.
Our focus is not to determine the best smoothing method,
which is more properly a subject for experiment or numerical analysis;
rather we show that a wide variety of methods can be
performed efficiently.

A commonly used technique, {\em Laplacian smoothing},
sweeps over the mesh, successively
moving each point to the centroid of its neighbors.
This technique lacks motivation
because it is not directly connected to
any specific mesh quality criterion; moreover, 
the result may not even remain a valid triangulation.  
But in practice Laplacian smoothing
spaces points evenly and gives two-dimensional
meshes of reasonable quality. 
In three dimensions, however, even spacing
does not guarantee good element quality.
A {\em sliver} tetrahedron is one that has
evenly spaced vertices, but very sharp angles;
for instance a sliver can be formed by slightly perturbing the vertices
of a square. (See~\cite{BCER} for a more detailed classification of
tetrahedra in terms of solid and dihedral angles.)  
Laplacian smoothing sometimes removes
slivers, but in large meshes it often leaves
clusters of slivers~\cite{FO}.

Freitag, Jones, and Plassmann~\cite{FJP,FJP2} proposed 
an alternative to Laplacian smoothing.
Rather than using the centroid, their optimization-based
method computes for each Steiner point a new placement
that maximizes the minimum angle in adjacent triangles. 
Freitag et al.{} use an iterative steepest-descent algorithm
to solve this optimal placement problem. 
Empirically this algorithm finds the optimum location
in an average of 2.5 steps, but Freitag et al.{}
do not prove their algorithm correct.

The same optimal placement problem was independently
considered by Matou\v{s}ek et
al.~\cite{MSW} as an instance of the paradigm
called generalized linear programming.
Matou\v{s}ek et al.{} show how to solve
this problem using an algorithm related to the dual simplex method.
(In retrospect, the steepest-descent algorithm
of Freitag et al.{} can be seen as a primal simplex method, but
its correctness is not directly justified by 
the work of Matou\v{s}ek et al.{}; correctness
follows from our analysis below.)

Minimum angle, however,
is not the only measure of mesh quality.
Various papers have provided theoretical justification for other
measures including maximum angle~\cite{BA}, maximum edge
length~\cite{SF}, minimum height~\cite{GCR}, minimum containing
circle~\cite{DS}, and---most recently---ratio of area to sum of 
squared edge lengths~\cite{BS}.
Data-dependent criteria~\cite{BS,DLR,R,RS} may be used in
adaptive meshing, which uses the finite
element method's output to improve the mesh for another run.

In this paper, we study optimization-based smoothing using
quality criteria such as those mentioned above. 
We show that, as in the case of minimum angle, many of these criteria 
give rise to {\em quasiconvex programs} and can be
solved by linear-time dual simplex methods
or steepest-descent primal simplex methods.
Because of the generality of these methods, they can
also solve mixed-criterion optimization problems.

We generalize the theory to quadrilateral meshes and to simplicial
meshes in three and higher dimensions. In these more complicated meshing
problems, effective smoothing methods are a more critical need and
asymptotic time complexity is more important. We show that again
quasiconvex programming often arises; for instance it can 
maximize the minimum solid angle.  
We believe optimization-based three-dimensional
mesh smoothing should outperform Laplacian smoothing in practice.
Indeed, in very recent experimental work
Freitag and Ollivier-Gooch~\cite{FO}
have shown that optimization-based smoothing for minimum dihedral angle 
outperforms Laplacian smoothing, both alone
and in conjunction with flipping.

Finally, we show that although several other optimal point placement
problems do not form quasiconvex programs, we
can solve them efficiently by other means.
This direction may also be relevant in practice;
Freitag and Ollivier-Gooch recommend smoothing for
the sine of the dihedral, a non-quasiconvex quality measure.

\section{Generalized Linear Programming}

Many problems in computational geometry, such as separating
points by a hyperplane, can be modeled directly as low dimensional
linear programs.
Many other problems, such as the circumcircle of a point
set, are not linear programs, but the same techniques often apply to
them. To explain this phenomenon, various authors have formulated a
theory of {\em generalized linear
programming}~\cite{A,G,MSW}.

A {\em generalized linear program} (GLP, also known as an {\em LP-type
problem}) consists of a finite set
$S$ of {\em constraints} and an {\em objective function} $f$ mapping
subsets of $S$ to some totally ordered space and satisfying the
following properties:
\begin{enumerate}
\item For any $A\subset B$, $f(A)\le f(B)$.
\item For any $A$, $p$, and $q$,
if $f(A)=f(A\union\{p\})=f(A\union\{q\})$, then
$f(A)=f(A\union\{p,q\})$.\footnote{Property~2 is often expressed in the
more complicated form that, if $A\subset B$ and $f(A)=f(B)$, then, for
any $p$,
$f(A)=f(A\union\{p\})$ iff $f(B)=f(B\union\{p\})$. A simple induction
shows this to be equivalent to our formulation.}
\end{enumerate}
The problem is to compute
$f(S)$ using only evaluations of $f$ on small subsets of~$S$.

For instance, in linear programming, $S$ is a
set of halfspaces and $f(S)$ is the point in the intersection of the
halfspaces at which some linear function takes its minimum value. Another
standard example of a GLP is the problem of computing the minimum radius
of a disk containing all of a set of
$n$ points; in this example, the finite set $S$ consists of the points
themselves, and $f(A)$ is the minimum disk.
It is not hard to see that this system satisfies the properties by which
a GLP was defined above: removing points can only make the radius shrink
or stay the same, and if a disk contains the additional
points $p$ and $q$ separately it contains them both together.

A {\em basis} of a GLP is a set $B$
such that for any $A\subsetneq B$, $f(A)<f(B)$.
The {\em dimension}~$d$ of a GLP is the maximum
cardinality of a basis.
With the standard example of the minimum disk problem,
the dimension turns out to be three, because each circle is determined by
two or three points.  This set of two or three points is the basis.

A number of efficient GLP algorithms are
known~\cite{AS,A,C,DF,G,MSW}. Their best running time is
$O(d\:n\kern0.05em T + f(d)E\log n)$ where $n$ is the number of
constraints,
$T$ measures the time to test a proposed solution against a
constraint (typically this is $O(d)$),
$f$ is exponential or subexponential, and $E$ is the time to find a
basis of a constant-sized subproblem.
Indeed, these algorithms are straightforward to implement and have small
constant factors, so they should be practical even for the modest values
of $n$ relevant in our problems. (The number of constraints should range
roughly from 10 to 100 in the planar problems, depending on how
complicated a criterion one chooses to optimize and on the degree of the
initial mesh, and may possibly reach several hundred in the
three-dimensional problems.)

Our GLPs have the
following form, which we call ``quasiconvex programming''.
We wish to minimize some objective function that is the pointwise maximum
of a finite set of functions.  Such a problem will be a
low-dimensional GLP if the level sets of the functions
(regions in which the function is bounded above by some particular value)
are all convex. Note that this does not necessarily imply that the
functions themselves are convex; in convex analysis, functions with
convex level sets are called {\em quasiconvex}.

More formally, define a {\em nested convex family} to be a map
$\kappa(t)$ from the nonnegative real numbers to compact convex sets in
$\R^d$ such that if
$a<b$ then
$\kappa(a)\subset\kappa(b)$, and such that
for all $t$, $\kappa(t)=\bigcap_{t'>t}\kappa(t')$.
Any nested convex family $\kappa$ determines
a function $f_\kappa(x) = \inf\,\{\,t \mathrel{|} x \in \kappa(t)\,\}$
on $\R^d$, with level sets that are the boundaries of $\kappa(t)$.
If $f_\kappa$ does not take a constant value on any open set,
and if $\kappa(t')$ is contained in the interior of
$\kappa(t)$ for any $t'<t$, we say that $\kappa$ is {\em continuously
shrinking}.

Note that, in our proof of Lemma~\ref{L:glp} below, we will consider the
restriction of convex families to affine subspaces; such restrictions
do not necessarily preserve the property of being continuously
shrinking.  However, if $\kappa$ is continuously shrinking, and its
restriction to any affine subspace $A$ has $f_\kappa=t$ on some open set
in $A$, then all points of this open set are on the boundary of
$\kappa(t)$ and $f_\kappa(t')$ must have empty intersection with $A$
for any $t'<t$.

\begin{lemma}\label{nonempty-inf}
Let $\kappa$ be a nested convex family, and let
$t^*=\inf\,\left\{\,t\mathrel{|}\kappa(t)\hbox{ is nonempty}\,\right\}$.
Then $\kappa(t^*)$ is nonempty.
\end{lemma}

\begin{proof}
Choose a point $p_i$ in the set $\kappa(t+1/i)$
for $i=0,1,2,\ldots$.  Since all of these points are contained in the
compact set $\kappa(t+1)$, they have a limit point $p^*$.
Then for any $i$, $\kappa(t+1/i)$ contains all but finitely many of the
points $p_i$, so $p^*$ is a limit point of the closed set
$\kappa(t+1/i)$ and must be in $\kappa(t+1/i)$.
Since $p^*$ is in all of the sets $\kappa(t+1/i)$ it is
in their intersection $\kappa(t^*)$.
\end{proof}

If $S=\{\kappa_1,\kappa_2,\ldots \kappa_n\}$ is a set of  nested convex
families, we define
$S(t)=\bigcap\left\{\,\kappa_i(t)\,\right\}$.
Then $S(t)$ is itself a nested convex family: each set $S(t)$ is the
intersection of closed bounded convex sets, hence is itself closed,
bounded, and convex.  The further requirement that
$S(t)=\bigcap_{t'>t}S(t')$ can easily be seen to follow by commutativity
of intersections.

If $S=\{\kappa_1,\kappa_2,\ldots \kappa_n\}$ is a set of  nested convex
families, and $A\subset S$, let
$$
f(A)=\inf\Big\{\,
		(t,x) \mathrel{\big|}
				x\in \mathop{\textstyle\bigcap}\limits_{\kappa_i\in A}\kappa_i(t)
\Big\}
$$
where the infimum is taken in the lexicographic ordering,
first by $t$ and then by the coordinates of~$x$.
Note that the values of $t$ are bounded below by zero
(because $\kappa_i(t)$ is only defined for nonnegative $t$),
so the infimum of $t$ exists.
The rest of this lexicographic infimum is also well defined
since Lemma~\ref{nonempty-inf}
shows that, if $t^*$ is the value determined by the infimum,
$A(t^*)$ is a nonempty compact set, and $x$ is simply the
lexicographic minimum of this set. We use this same lexicographic ordering
to compare the values of
$f$ on different subsets of~$S$.

Recall Helly's theorem (e.g., see \cite{A}):
If a family of compact convex sets in $\R^d$
(or a finite family of non-compact convex sets)
has an empty intersection, then some $(d+1)$-tuple of those sets also
has an empty intrsection.

We define a {\em quasiconvex program}
to be a finite set $S$ of nested convex families,
with the objective function $f$ described above.

\begin{lemma}\label{L:glp}
Any quasiconvex program forms a GLP of dimension at most $2d+1$.
If each $\kappa_i$ in the set $S$ is either
constant or continuously shrinking, the dimension is at most $d+1$.
\end{lemma}

\begin{proof}
Property~1 of GLPs is obvious.
Property~2 follows from the observation that, if
$(t^*,x^*)=f(A)$, then $f(A)=f(A\union\{\kappa_j\})$
if and only if $x^*\in\kappa_j(t^*)$.
It remains only to show the stated bounds on the dimension.

First consider the general case, where we do not assume continuous
shrinking of the families in $S$. Let $(t^*,x^*)=f(S)$.
For any $t<t^*$, $S(t)=\bigcap\kappa_i(t)=\emptyset$
so by Helly's theorem some $(d+1)$-tuple of sets $\kappa_i(t)$ has empty
intersection.  Since there are only finitely many $(d+1)$-tuples,
we can choose a tuple $B^{-}$ that has an empty intersection for
all $t<t^*$.  Then $f(B^{-})=(t^*,x)$ for some $x$, so the presence of
$B^{-}$ forces the GLP solution to have the correct value of $t$.
By Lemma~\ref{nonempty-inf}, $S(t^*)\neq\emptyset$, so
$x^*$ is the minimal point in $S(t^*)$, and is determined by some
$d$-tuple $B^{+}$ of the sets $\kappa_i(t^*)$.  Then $f(B^{-}\cup
B^{+})=f(S)$, so some basis of $S$ is a subset of $B^{-}\cup B^{+}$ and
has cardinality at most $2d+1$.

Finally, suppose  each $\kappa_i$ in $S$ is
constant or continuously shrinking.  Our strategy will be to again find a
tuple
$B^{-}$ that determines $t^*$, and a tuple $B^{+}$ that
determines $x^*$, but we will use continuity to make the sizes of
these two tuples add to at most $d+1$.

$S(t^*)$ has empty interior: otherwise, we could find an open region $X$
within
$S(t^*)$, and a family $\kappa_i$ such that $\kappa_i(t)\cap
X=\emptyset$ for any $t<t^*$, violating the assumption that $\kappa_i$
is constant or continuously shrinking.
If the interior of some $\kappa_i(t^*)$ contains a point of
the affine hull of $S(t^*)$, we say that $\kappa_i$ is ``slack'';
otherwise we say that
$\kappa_i$ is ``tight''.  The boundary of a slack $\kappa_i$
intersects
$S(t^*)$ in a subset of measure zero (relative to the affine hull of
$S(t^*)$), so we can find a value
$x$ in the relative interior of
$S(t^*)$ and not on the boundary of any slack $\kappa_i$. Form the
projection
$\pi:\R^d\mapsto\R^{d-\dim S(t^*)}$ perpendicular to $S(t^*)$.

For any ray $r$ in $\R^{d-\dim S(t^*)}$ starting at the point
$\pi(S(t^*))$, we can lift that ray to a ray $\hat r$ in $\R^d$
starting at $x$, and find a hyperplane
containing $S(t^*)$ and separating the interior of some
$\kappa_i(t^*)$ from $\hat r\setminus\{x\}$.
This separated $\kappa_i$ must be
tight (because it has $x$ on its boundary as the origin of the ray)
so the separating hyperplane must contain the affine hull of $S(t^*)$
(otherwise some point in $S(t^*)$ within a small neighborhood of $x$
would be interior to $\kappa_i$).  Therefore the hyperplane
is projected by $\pi$ to a lower dimensional hyperplane separating
$\pi(\kappa_i(t^*))$ from $\pi(S(t^*))$.
Since one can find such a separation for any ray,
$\bigcap_{\hbox{tight }\kappa_i}\pi(\kappa_i(t^*))$ can not contain any
points of any such ray and must consist of the single point
$\pi(S(t^*))$.

At least one tight $\kappa_j$
must
be continuously shrinking (rather than constant), since otherwise 
$S(t)$ would be nonempty for some $t<t^*$.
The intersection of the interior of $\pi(\kappa_j(t^*))$ with
the remaining projected tight constraints $\pi(\kappa_i(t^*))$ is empty,
so by
Helly's theorem, we can find a $(d-\dim S(t^*) + 1)$-tuple
$B^{-}$ of these convex sets having empty intersection, and the
presence of $B^{-}$ forces the GLP solution to have the correct value of
$t$. Similarly, we can reduce the size of the set $B^{+}$ determining
$x^*$ from $d$ to $\dim S(t^*)$, so the total size of a basis
is at most $(d-\dim S(t^*) + 1)+\dim S(t^*)=d+1$.
\end{proof}

The first part of this lemma is similar to~\cite[Theorem~8.1]{A}.
Note that we only used the assumption of convexity to prove the
dimension bound; similar nested families of non-convex sets still
produce GLP problems, but could have arbitrarily large dimension.

By Lemma~\ref{L:glp} we can solve quasiconvex programs using GLP
algorithms.  We can also perform a more direct local optimization
procedure to find $(t,x)$: since $S(t)$
is a nested convex family we can find $f(S)$
by applying steepest descent,
nested binary search, or other local optimization techniques
to find the point minimizing the associated function $f_S(x)$.  Thus we
can justify the correctness of the local optimization mesh smoothing
procedure used by Freitag et al.  In practice, it may be appropriate to
combine this approach with the dual simplex methods coming from GLP
theory by using steepest descent to perform the basis
exchange operations needed in GLP algorithms.

\section{Quasiconvex Mesh Smoothing in $\R^2$}

\begin{figure*}[t]
$$\efig{2in}{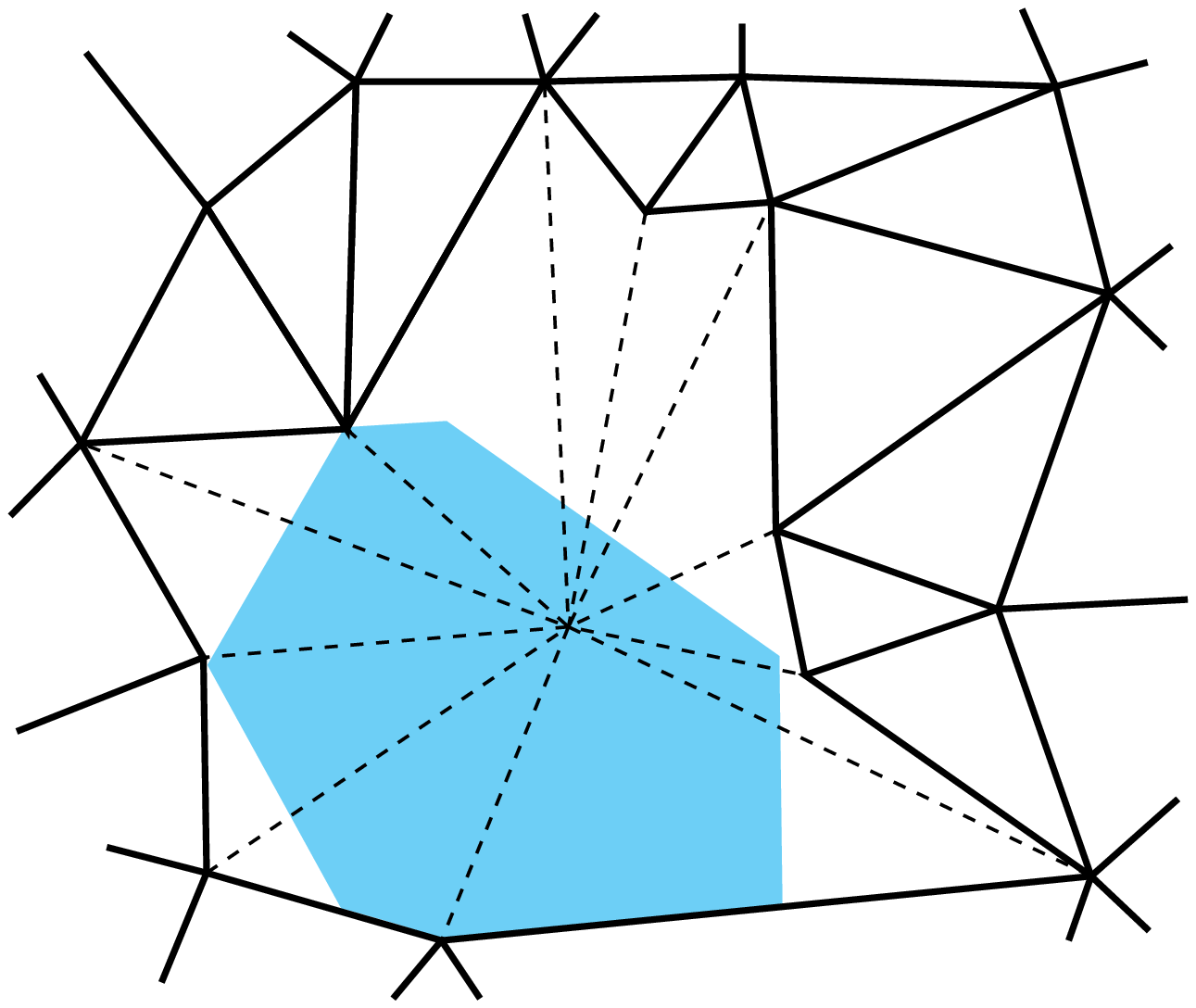}\qquad\efig{1.5in}{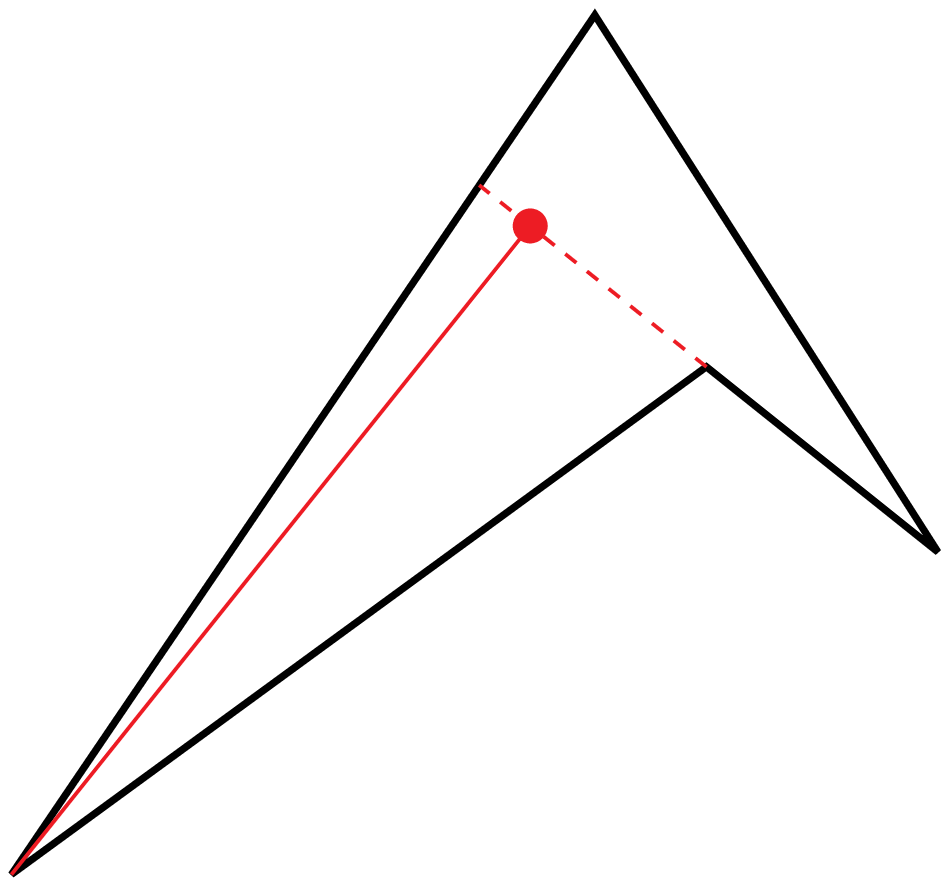}$$
\caption{(a) Steiner point may move within kernel of star-shaped region
formed by its removal; (b) For size-based criteria such as length the
optimal placement may be on the kernel boundary.}
\label{F:ker}
\end{figure*}

Let $q(\Delta)$ measure the quality of a triangulation
element $\Delta$.  We are given a triangulation, and wish to move
one of its Steiner points in such a way as to minimize $\max q(\Delta_i)$,
where the maximization occurs over elements incident to the moving point.

In this section we describe ways of formulating
such problems as quasiconvex programs.  We can assume without loss
of generality (e.g. by appropriate change of variables) that
$q(\Delta)\ge 0$ for any $\Delta$. The basic idea is to construct for each
$\Delta_i$ a nested convex family
$\kappa_i(t)=\{\,x\mid q(\Delta_i(x))\le t\,\}$, where $\Delta_i(x)$
indicates the triangle formed by moving the Steiner point to position
$x$.  In other words, if we are given a bound
$t$ on the triangulation quality, $\kappa_i(t)$ is the {\em feasible
region} in which placement of the Steiner point will allow $\Delta_i$ to
meet the quality bound.  Finding the optimal Steiner
point placement is equivalent to finding the optimal quality
bound that allows a feasible placement.

The families $\kappa_i(t)$ are clearly nested and closed, and they
satisfy the intersection property used in the definition of nested
convex families, but they may not be convex or bounded.
Convexity will need to be proven using the detailed
properties of the quality measure
$q$.
Continuous shrinking may or may not hold depending on the quality measure
$q$.  Boundedness can be imposed (while preserving continuous shrinking)
by intersecting
$\kappa_i(t)$ with the set of points within distance $\exp(t)$ of a
bounding ball of the triangulation.

One can then find the optimal placement $x$ by solving the
quasiconvex program associated with this collection of nested convex
families. To make sure that the result is a valid triangulation,
we add additional halfspace constraints to our collection,
forming constant nested families, to
force $x$ into the kernel of the star-shaped polygon
formed by removing the Steiner point from the
triangulation (Figure~\ref{F:ker}(a)).

\begin{figure*}[t]
$$\efig{4in}{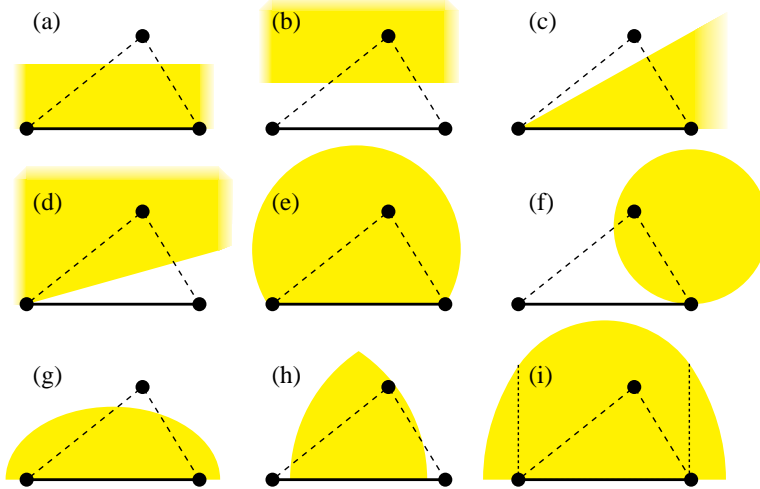}$$
\caption{Feasible regions for planar mesh smoothing quality criteria:
(a) minimizing maximum area or external altitude;
(b) maximizing minimum area, external altitude, or external aspect ratio;
(c) minimizing maximum external angle;
(d) maximizing minimum external angle, or maximizing
minimum internal altitude;
(e) maximizing minimum internal angle;
(f) maximizing internal aspect ratio;
(g) minimizing maximum perimeter;
(h) minimizing maximum edge length
(a similar but larger lune occurs when minimizing diameter);
(i) minimizing containing circle.}
\label{F:feas}
\end{figure*}

It remains to show convexity of the feasible regions $\kappa_i(t)$ for
various quality measures.  In the remainder of this section, we
describe these measures and their corresponding feasible regions.  As
shown in Figure~\ref{F:feas}, many different criteria have identical
feasible regions; however they do not necessarily lead to the same
Steiner point placement as the parametrization of the nested families
could differ.

\begin{description}
\item[Area.]
The feasible regions for
maximizing minimum triangle area
are strips parallel to the fixed
(external) sides of the triangles.  In the presence of the halfspace
constraints forcing the Steiner point into the kernel of its
polygon, we can simplify these strips to halfspaces. 
The intersection of one such halfspace and the corresponding kernel
constraint is shown in Figure~\ref{F:feas}(a).  One can also maximize
minimum area, using a halfspace with the same boundary but opposite
orientation (Figure~\ref{F:feas}(b)).

\item[Altitude.]
The {\em external altitude} of $\Delta_i$ (the altitude having
the fixed side of $\Delta_i$ as its base) can be minimized or
maximized using halfspace feasible regions identical to those for area
(Figure~\ref{F:feas}(a,b)). The feasible regions in which the
other two altitudes are at least $h$ are
the intersections of pairs of halfspaces through one fixed point, passing
at distance $h$ from the other point; one such
halfspace is shown in Figure~\ref{F:feas}(d) and the other is its
vertical reflection.  The feasible regions for minimizing the maximum
internal altitude are not convex.

\item[Angle.]
As noted by Matou\v{s}ek et al.~\cite{MSW},
one can maximize the minimum angle by using constraints of two types.
For the internal angles at the Steiner points, the region in which
the angle is at least $\theta$ forms either the union
or intersection of two congruent circles (as $\theta$ is acute
or obtuse respectively) having the fixed side of $\Delta_i$ as a
chord.  In the former case this may not be convex, but in the presence of
the kernel constraints we can simplify the feasible region
to circles (Figure~\ref{F:feas}(e)).  The regions in which the external
angles are at least
$\theta$ form wedges bounded by rays through a fixed vertex of
$\Delta_i$, which can again be simplified in the presence of the kernel
constraints to halfspaces (Figure~\ref{F:feas}(d)).
It is also natural to minimize the maximum angle;
unfortunately the feasible regions for the internal angles are
non-convex (complements of circles).
However one can still minimize the maximum angle at external
vertices, using halfspace regions (Figures~\ref{F:feas}(c)).

\item[Edge length.]
The feasible region for minimizing the length of the internal edges of
$\Delta_i$ is an intersection of two circles of the given radius,
centered on the fixed vertices of $\Delta_i$ (Figure~\ref{F:feas}(h)).
We can use the same two-circle constraints (with
larger radii than depicted in the figure) to minimize the
maximum element diameter.

\item[Aspect ratio.]
The {\em aspect ratio} of a triangle is the ratio of its longest side
length to its shortest altitude.  We consider separately the ratios of the
three sides to their corresponding altitudes; the maximum of these three
will give the overall aspect ratio.  The ratio of external sides to
altitude has a feasible region (after taking into account the
kernel constraints) forming a halfspace parallel to the external side,
like that in Figure~\ref{F:feas}(b).  To determine the aspect ratio on
one of the other two sides of a triangle $\Delta_i$, normalize the
triangle coordinates so that the moving point has coordinates $(x,y)$ and
the other two have coordinates $(0,0)$ and $(1,0)$.  The side length is
then $\sqrt{x^2+y^2}$, and the altitude is $y/\sqrt{x^2+y^2}$,
so the overall aspect ratio has the simple formula $(x^2+y^2)/y$.
The locus of points for which this is a constant $b$
is given by $x^2+y^2=b\:y$, or equivalently
$x^2+(y-(b/2))^2=(b/2)^2$.  Thus the feasible region is a
circle tangent to the fixed side of $\Delta_i$ at one of its two
endpoints (Figure~\ref{F:feas}(f)).

\item[Perimeter.]
The feasible region for minimizing the maximum perimeter is an
ellipse (Figure~\ref{F:feas}(g)).

\item[Circumradius and containing circle.]
The feasible regions for minimizing the maximum circumradius are nonconvex
lunes bounded by pairs of circular arcs. However, minimizing the maximum
{\em containing circle} (the smallest circle containing the given
triangle, without necessarily having the vertices on its boundary)
produces convex feasible regions, formed by using the same region as the
circumcircle within a vertical slab perpendicular to the fixed segment of
the triangle, and a lune similar to that for edge length or diameter
outside the slab.  These regions' boundaries are three circular
arcs, meeting at common tangents, with the radius of the middle arc equal
to half that of the arcs on either side (Figure~\ref{F:feas}(i)).

\item[Inradius.]
The feasible region for maximizing the minimum inradius of any triangle
can be found as follows. Assume without loss of generality that the
two fixed points have coordinates $(0,0)$ and $(0,1)$, the moving point
has coordinates $(x,y)$, and the inradius bound is $r$. We can
then place the incenter at a point
$(a,r)$ and solve simultaneous equations stating that lines from $(0,0)$
to
$(x,y)$ and from $(1,0)$ to $(x,y)$ are at distance r from this
point.  The solution to these equations was simplified in {\em
Mathematica} to
$$-8 r^3 x + 8 r^3 x^2 + 4 r^2 y - 4 r^4 y + 4 r^2 x\: y 
- 4 r^2 x^2 y - 4 r\: y^2 + 8 r^3 y^2 + y^3 - 4 r^2 y^3
= 0.$$
Affine transformation of the coordinates further simplifies this to
$$-8r^5 + r^2 y - 20r^4 y - x^2 y + 2 r\: y^2 - 
16 r^3 y^2 + y^3 - 4 r^2 y^3 = 0,$$
which has only one term involving
$x$, letting us solve this as $x=f(y)$ for a function $f$ in the form of
the square root of a rational function:
$$x=\pm f(y)=\pm \sqrt{(r+y)^2(y(1-4r^2)-8r^3)/y}.$$
To show that this bounds a convex region, we need only show that $f$ has
nonpositive second derivative within the range of values $y$ leading to
a feasible solution.  We used {\em Mathematica} to compute this
derivative:
$$f''(y)={8 r^4 (r+y)^3 (6r^3 - y + 2 r^2 y)
\over y^{5/2} (r+y)^3 (y(1-4r^2)-8r^3)^{3/2}}.$$
Most of the terms in this formula clearly have a consistent sign.
The final polynomial in the denominator has a root at
$y=8r^3/(1-4r^2)$, which turns out to be the point at which $y$ is
minimum, corresponding (in the original coordinate system prior to our
affine transformation) to
$x=1/2$; smaller values of $y$ are infeasible.  The final polynomial in
the numerator has a root at $y=6r^3/(1-2r^2)$, which is always below this
minimum feasible value of $y$.  Therefore $f''$ has a consistent sign
throughout the interval of interest, and
the feasible region for inradius is convex.

\item[Area over squared edge length.]
Bank and Smith~\cite{BS} define yet another measure of the quality of a
triangle, computed by dividing the triangle's area by the sum of the
squares of its edge lengths.  This gives a dimensionless quantity which
Bank and Smith normalize to be one for the equilateral triangle (and
less than one for any other triangle).  They then use this quality
measure as the basis for a local improvement method for mesh smoothing.
As Bank and Smith show, the feasible region for this measure
forms a circle centered on the perpendicular bisector of the two fixed
points, so our methods offer an alternative way to find the
optimum point placement.

\item[Mixtures of criteria.]
We have described the various optimization criteria above as if
only one is to be used in the actual mesh smoothing algorithm.  But
clearly, the same formulation applies to problems in which we
combine various criteria, for instance some measuring element shape and
others measuring element size, with the overall quality of an element
equal to the weighted maximum of these criteria.  Indeed, this idea can
alleviate a problem with 
criteria such as edge length,
perimeter, etc., which depend more strongly on the size of an
element than on its shape: if one optimizes such a criterion on its own,
the optimal point placement may lie on the boundary of the kernel,
giving rise to a degenerate triangulation (Figure~\ref{F:ker}(b)).
If one combines these criteria with scale-invariant criteria such as
angles or aspect ratio,
this complication cannot occur.  We define the quality of a mixture of
criteria $q_i$ to be $\max w_i q_i$, where the weights $w_i$ may
be chosen arbitrarily.  (Even more generally we could replace the linear
function $w_i q_i$ with any monotonic function of $q_i$.)
To solve such a mixed problem, we simply include
constraints for each different criterion in the combination.
\end{description}

\begin{theorem}
The Steiner point placement optimizing the criteria described
above, or a weighted maximum of criteria, can be computed
in linear time by quasiconvex
programming.
\end{theorem}

\begin{proof}
By Lemma~\ref{L:glp} we can solve these problems using any algorithm for
GLP-type problems.  As noted earlier, a number of algorithms are known
for solving such problems in a linear number of operations, where each
operation involves 
testing a potential solution against one of the constraints
(which in our case amounts to computing the quality of a single element)
or finding the solution of a subproblem of constant size.
These constant-size subproblems can be solved in constant time
in the algebraic decision tree model standard for geometric algorithms.
\end{proof}

\section{Quadrilateral Mesh Smoothing}

Much of the same theory we have outlined above applies equally well to
{\em quadrilateral meshes}, meshes consisting of planar
straight-line graphs in which all faces are convex quadrilaterals.
In this case, to preserve element convexity, the Steiner point must not
only stay within the kernel of the star-shaped polygon formed by adjacent
elements, it must also avoid crossing any element diagonal.
Also, some of the quality measures outlined above do not make as much
sense when applied to quadrilaterals, and others have feasible regions
differing somewhat from those for triangular elements. We outline below
some possible quality criteria for quadrilateral meshes and the changes
needed to adapt our triangular-mesh smoothing methods to these criteria.

\begin{description}
\item[Area, angle, edge length, perimeter.]
The feasible regions for placing a Steiner point according to these
criteria are essentially the same as for triangular meshes.

\item[Width.]
This corresponds to the altitude of a triangle. The width is the minimum
distance between a point and one of the two opposite edges, and the
minimum width can be maximized by a feasible region formed by the
intersection of six halfspaces, one for each vertex-edge pair involving
the moving point.

\item[Containing circle.]
The minimum containing circle for a quadrilateral is the same as the
largest of the four circles formed by choosing three of the four points
(in each of four possible ways)
and considering the containing circle of that triple of points.
Therefore, the feasible regions for minimizing the maximum
containing circle are the intersections of three of the regions arising
in the triangular case, one for each of the three triples involving the
moving point.  Since each of these regions is convex, the overall
feasible region is convex.

\item[Diameter.]
The diameter of a quadrilateral is either its longest edge or its longest
diagonal. Hence the feasible region for
diameter is an intersection of circles, similar to that for edge length,
but with the difference that we include a third circle centered on the
vertex opposite the moving point.

\item[Inradius.]
Our proof that the triangle inradius function has convex feasible
regions does not immediately generalize to quadrilaterals.
We conjecture that quadrilateral inradii also give convex feasible
regions.
\end{description}

\begin{theorem}
The Steiner point placement optimizing the quadrilateral mesh criteria
described above (except possibly inradius), or a weighted maximum of
criteria, can be computed in linear time by quasiconvex programming.
\end{theorem}

Some other natural quality measures for quadrilaterals, such as cross
ratio (ratio of products of opposite side lengths) and sums of opposite
pairs of angles, do not have convex feasible regions,
but (since their feasible regions are bounded by circular arcs)
can be optimized using the techniques described below in
Theorem~\ref{minmaxangle}.

\section{Mesh Smoothing in Higher Dimensions}

Many of the two-dimensional quality criteria discussed above
have higher-dimensional generalizations that also have convex feasible
regions.

\begin{description}
\item[Volume and altitude.]
Just as the area of a triangle with a fixed base is proportional to its
height, the volume of a simplex with a fixed base is proportional to its
altitude.  The triangulation minimizing the maximum volume, or
maximizing the minimum volume, can be found using feasible regions in
the form of halfspaces parallel to the fixed face of the simplex.
The same type of feasible region can be used to optimize the altitude at
the moving Steiner point.  The feasible regions for maximizing the
minimum of the other altitudes are the intersections of pairs of
halfspaces through $d-1$ of the fixed points.

\item[Boundary measure.]
The measure of any boundary face of a simplex is proportional to the
distance of the moving Steiner point from the affine hull of the
remaining fixed points on the facet, so one can minimize the maximum
face measure using ``generalized cylinders'' formed by taking a
cartesian product of a sphere with this affine hull.  In particular the
Steiner point placement minimizing the maximum edge length can be found
by using spherical feasible regions centered on each fixed point,
and in $\R^3$ the placement minimizing the maximum triangle area can be
found using cylindrical feasible regions centered on each fixed edge.
These face measures are convex functions, so their sums are also convex,
implying that the level sets for total surface area of all triangles in a
tetrahedron, or total length of all edges in a tetrahedron, again form
convex feasible regions.

\item[Containing sphere.]
As in $\R^2$, the feasible regions for
the minimum containing sphere are bounded by $2^{d}-1$ algebraic
patches, in which the containing sphere has some
fixed set of vertices on its boundary.  These patches
meet the plane of the fixed vertices perpendicularly, and
are locally convex (they are figures of rotation of lower
dimensional feasible regions, except for the one corresponding to the
region in which the containing sphere equals the circumsphere, which is
a portion of that sphere).  In $\R^3$, these patches are portions of
spheres and tori.  Further, they meet at a continuous boundary (since
the containing sphere radius is a continuous function of the moving
point's location) and are continuously differentiable where they meet
(at each point where two patches meet, they share tangent planes with the
containing sphere itself).  Thus these patches combine to form a convex
region.

\item[Dihedrals.]
The dihedral angles of a simplex are formed where two faces
meet along an {\em axis} determined by some $d-1$ points.
If these axis points are all fixed, one of the two faces is
itself fixed, and the feasible region is a halfspace forming the
given angle with this fixed face.  However, if the axis includes the
moving point, the feasible regions are in general non-convex.

\item[Solid Angles.]
As we show in the next section, the feasible regions for maximizing the
minimum solid angle (measured at the fixed points of each tetrahedron, for
three-dimensional problems, or at the moving point in any dimension) are
convex.
\end{description}

\begin{theorem}
In any constant dimension, the Steiner point
placement optimizing each of the criteria described above except
exterior solid angle, or a weighted maximum of criteria, can be
computed in linear time by quasiconvex programming.
The exterior solid angles as well can be optimized in three dimensions.
\end{theorem}

\section{Feasible Regions for Solid Angles}

We now prove that the feasible regions for maximizing the minimum solid
angles of the mesh elements are convex, for the angles at the moving
point, in any dimension, and  for the angles at fixed points of tetrahedra
in $\R^3$ only. Convexity of the feasible regions for solid angles at
fixed points in higher dimensions remains open.

We start with the simpler case, in which we are interested in the solid
angle at one of the fixed vertices of a tetrahedron in $\R^3$.
This angle can be measured by projecting the other three vertices
onto a unit sphere centered on the fixed vertex, and measuring the area
of the spherical triangle formed by these three projected points.
If the three projected points are represented by three-dimensional unit
vectors $a$, $b$, and $c$ (with $a$ representing the
moving point and
$b$, $c$, and the origin representing the three fixed points)
then the solid angle $E$ at the origin satisfies the equation
$$\tan(E/2)=
{a\cdot(b\times c)
\over
1 + b\cdot c + c\cdot a + a\cdot b}$$
\cite{F}.
Therefore, the boundary of the feasible region (on the unit sphere)
is given by an equation of the form
$$a\cdot(b\times c)
=k(1 + b\cdot c + c\cdot a + a\cdot b),$$
which is linear in $a$ and therefore forms a circle on the unit
sphere.  (Note that unlike in the planar case, this circle does not pass
through $b$ and $c$, but instead passes through their diametric
opposites.)  In terms of the original unprojected
points, the feasible region is therefore a convex circular cone.

To prove that the feasible regions for the interior solid angles are
also convex, we use some facts from convex analysis~\cite{DJ}.
A function $f(v)$ from some convex subset of a vector space $V$ to
$\R$ is said to be {\em convex} if, for any $x,y\in V$, and any $0\le
t\le 1$,
$$f(t\cdot x + (1-t)\cdot y)\le t\cdot f(x) + (1-t)\cdot f(y).$$
A function $f(v)$ is said to be {\em quasiconcave} if its level sets
$\{v\mid f(v)\ge k\}$ are convex.  A function is $s$-concave if $f(v)^s$
is convex; in the cases of interest to
us $s$ will always be negative.  If $f$ is quasiconcave we also say that
it is $(-\infty)$-concave (and if $f$ is logconcave, i.e. if $\log f$ is
convex, we also say that it is 0-concave).
%%  We allow functions to take infinite
%% values; these can be interpreted by the squeamish as shorthand for
%% appropriate limits.

%% \begin{lemma}\label{prod-conc}
%% Let $f(u)$ be $s$-concave for $s\le 0$, and $\kappa$ be a convex set.
%% Then the function $h(u,v)=f(u)$ is $s$-concave on the cartesian
%% product of the domain of $f$ with $\kappa$.
%% \end{lemma}

%% \begin{proof}
%% Let $U$ denote the domain of $f$, and $V$ denote the domain of $g$.
%% Let $(x,y)\in U\times V$.  If $x$ or $y$ projects to a point
%% in $V$ outside $\kappa$, $h(x)^s$ or $h(y)^s$ will be $+\infty$ and the
%% inequality defining the convexity of $h$ will be trivially satisfied.
%% Otherwise the values of $h^s(x)$ and $h^s(y)$ come from the projections
%% of $x$ and $y$ into $U$, and convexity of $h^s$ follows from the
%% assumed convexity of $f^s$.
%% \end{proof}

The next result appears as \cite[Theorem 3.21]{DJ}.
The ``usual conventions'' from that source imply that, if $s=-1/n$, the
integral is $(-\infty)$-concave i.e. quasiconcave.

\begin{lemma}\label{int-conc}
Let $f$ be $s$-concave on an open convex set $C$ in $\R^{m+n}$.
Let $C^*$ be the projection of $C$ on $\R^m$ and for $x\in C^*$,
let $C(x)$ be the $x$-section of $C$.  Define
$$f^*(x)=\int_{C(x)} f(x,y)\,d\kern 0.05em y,\quad x\in C^*.$$
If $-1/n\le s\le \infty$, then $f^*$ is $s^*$-concave on $C^*$, where
$s^*=s/(1+n\:s)$ with the usual conventions when $s=-1/n$ or $s=\infty$.
\end{lemma}

\begin{corollary}\label{convolve}
Let $f:U\mapsto\R$ be $(-1/k)$-concave, and let $g:V\mapsto\{0,1\}$ be the
characteristic function of a convex set $\kappa$
in a $k$-dimen\-sion\-al subspace $V$ of $U$. Then the convolution of $f$
and $g$ is quasiconcave.
\end{corollary}

\begin{proof}
Let $h(u,v)=f(u)$, defined on the cartesian product of $U$ with $V$.
Then $h$ is also $(-1/k)$-concave, and the convolution can be computed as
$h^*(u-v)$. The result follows from Lemma~\ref{int-conc}.
\end{proof}

A special case of Corollary~\ref{convolve}, in which $k$ equals the
dimension $d$ of the domain of $f$, appears (with a different proof) as
\cite[Theorem 3.24]{DJ}.  For our application, we are interested in a
different case, in which $k=d-1$.
The solid angle of a $d$-simplex in $d$-dimensional
space, measured at the moving point, can be interpreted as the fraction
of the field of view at that moving point taken up by the convex hull
$\kappa$ of the remaining fixed points.  This fraction can be computed
as the convolution of the characteristic function of $\kappa$
with a function $f(v)$ measuring the fraction of field of view taken by
an infinitesimally small surface patch of $\kappa$.
This function $f(v)$ is inversely proportional to the square
($d-1$ power, for general $d$) of the distance from $v$ to the
patch, and directly proportional to the sine of the incidence angle of
$v$ onto the patch. If we translate this patch to the origin, $f$ has the
simple form $(v\cdot e)/|v|^d$ where $e$ is a vector normal to the patch.

\begin{lemma}\label{s-concave}
The function $f(v)=(v\cdot e)/|v|^d$, defined
on the open halfspace $v\cdot e>0$, is $-1/(d-1)$-concave.
\end{lemma}

\begin{proof}
Because of the rotational symmetry of $f$, we need only prove this
for the two-dimensional function $f(x,y)=y/(x^2+y^2)^{d/2}$
in the halfplane $y>0$.
We used {\it Mathematica} to compute the principal determinants of the
Hessian of $f^s$.  These are
$$
{\partial^2\over\partial y^2} f(x,y)^{-1/(d-1)} =
{ d\, x^2 y^{1/(d-1)} (x^2 + y^2)^{d/(2d-2)}(x^2 + (d-1) y^2)
\over (d-1)^2 y^2 (x^2+y^2)^2}
$$
which is always positive (for $y>0$, $d>1$), and
$$
\left({\partial^2\over\partial x^2}
{\partial^2\over\partial y^2}
-{\partial^2\over\partial x\partial y}
{\partial^2\over\partial y\partial x}\right)f(x,y)^{-1/(d-1)}
=0.
$$
Since both principal determinants are non-negative,
the function is convex.
\end{proof}

\begin{theorem}
The feasible region for the solid angle at the moving point of a
simplex is convex.
\end{theorem}

\begin{proof}
As described above, we can express the solid angle as the convolution of
$f(v)$ with the characteristic function of the convex hull of the fixed
points.  By Lemma~\ref{s-concave}, $f$ is $-1/(d-1)$-concave within a
halfspace defined by the kernel constraints.  Therefore we can use
Corollary~\ref{convolve} to show that the solid angle is quasiconcave
and therefore has convex level sets.
\end{proof}

Our proof for the interior solid angles generalizes to any dimension,
but that for the exterior solid angles does not.
There seems to be some correspondence between the feasible regions of
interior solid angles in dimension $d$, and the feasible regions of
exterior solid angles in dimension $d+1$; perhaps this correspondence
can be exploited to show that the exterior solid angle feasible regions
are convex in higher dimensions as well.

\section{Non-quasiconvex Mesh Smoothing}

We have seen that many mesh smoothing criteria give
rise to quasiconvex programming problems; however, 
other criteria, including minmax angle, minmax circumradius, and maxmin
perimeter, do not have convex feasible regions.

Perhaps this can be seen as evidence that these measures are less
appropriate for mesh smoothing applications, since it means among
other things that there may be many local optima instead of one global
optimum.  Indeed, it seems likely that the height and perimeter criteria
mentioned above do not lead to good element shapes.
However there is evidence that the maximum angle is
an appropriate quality measure for finite element meshes~\cite{BA},
so we now discuss methods for optimizing this measure.
Our results should be seen as preliminary and unready for
practical implementation.

\begin{theorem}\label{minmaxangle}
We can find the placement of a Steiner point in a star-shaped polygon,
minimizing the maximum angle, in time $O(n\log^c n)$ for some constant
$c$.
\end{theorem}

\begin{proof}
Each feasible region in which some
particular angle is at most $\theta$ forms either a halfplane or
the complement of a disk. The lifting transformation
$(x,y)\mapsto(x,y,x^2+y^2)$ maps these regions to halfspaces in $\R^3$;
$\theta$ is feasible if the intersection of all these halfspaces meets
the paraboloid
$z=x^2+y^2$.  The result follows by applying parametric
search~\cite{Meg} to a parallel algorithm that constructs the
intersection~\cite{AGR,Good} and tests whether any of its features crosses
the paraboloid.
\end{proof}

We can of course combine the maximum angle with the many
other criteria, including circumradius, for which the feasible regions are
bounded by lines and circles.

An alternate approach suggests itself, which may have a better chance of
leading to a practical algorithm.  Define a generalized Voronoi diagram
the cells of which determine which mesh angle would be worst if the
Steiner point were placed in the cell.  Are the cells of this diagram
connected? If so it seems likely that generalized Voronoi diagram
algorithms~\cite{KL,KMM,MMO} can construct this
diagram in time $O(n\log n)$ or perhaps even $O(n)$.
We could then find the optimal placement by examining
the features of this diagram.

Finally, we consider one last criterion, minimum total edge length.
This does not fit into our quasiconvex programming framework,
since the overall quality is a sum of terms from each element rather than
a minimum or maximum of such terms; however the corresponding optimal
triangulation problem remains a topic of considerable theoretical
interest~\cite{DM,LK}. A mesh improvement phase might also help reduce
the (large) constant factors in known approximate minimum weight Steiner
triangulation algorithms~\cite{E}.
Without the kernel constraints enforcing that the result is a valid
triangulation, the problem of placing one Steiner point to minimize the
total distance to all other points is a facility location problem known
as the {\em Weber} or {\em Fermat-Weber problem}.  Although it has no
good exact solution (the solution point is a high degree polynomial in the
inputs~\cite{B,CM}) one can easily solve it
approximately by steepest descent~\cite{W}. The kernel constraints do not
change the overall nature of this solution. Thus this version of the mesh
smoothing problem can again be solved efficiently.

\section{Conclusions}

We have described a general framework for theoretical analysis of mesh
smoothing problems, and have shown how to perform optimal Steiner point
placement efficiently for many important quality
measures.  There remain some open problems, for instance it is not clear
to what extent our results extend to hexahedral meshing (in which one
cannot generally move a single vertex at a time while preserving
element convexity).
There also remain some quality measures that may possibly be quasiconvex,
but for which a proof of quasiconvexity has eluded us. However we
believe the most important directions for future research are empirical:
which of the criteria we have described leads to the best quality meshes,
and to what extent can theoretical generalized linear programming
algorithms serve as practical methods for the solution of these problems?

\end{document}